\documentclass[11pt,a4paper]{article}
\pdfoutput=1
\usepackage{jcapmod}
\usepackage{graphicx}
\usepackage{amsfonts}
\usepackage{subfigure}
\usepackage{hyperref}
\usepackage{color}        
\usepackage{amsmath}
\usepackage{empheq}
\usepackage{amssymb}
\usepackage{tikz}
\usepackage{tabularx}
\usepackage{booktabs}
\usepackage{mdframed}
\usepackage{lipsum}

\newcommand{\de}{\partial}
\newcommand{\gaunt}{\mathcal{G}^{m_1 m_2 m_3}_{\ell_1 \ell_2 \ell_3}}

\newcommand{\Ys}[1]{Y^\ast_{\ell_{#1} m_{#1}} (\hat{n}_{#1})}

\newcommand{\ys}[1]{Y^\ast_{\ell_{#1} m_{#1}} (\hat{n})}
\newcommand{\n}{\hat{n}}
\newcommand{\alm}[1]{a_{\ell_{#1} m_{#1}}(z_{#1})}
\newcommand{\Deltag}[1]{\Delta_{\rm g}(z_{#1}, \n_{#1})}
\def\d{{\rm d}}
\def\be{\begin{equation}}
\def\ee{\end{equation}}
\def\bea{\begin{eqnarray}}
\def\eea{\end{eqnarray}}
\def\ba{\begin{align}}

\def\bi{\begin{itemize}}
\def\ei{\end{itemize}}
\begin{document}

\vspace{2cm}
\begin{center}
{\fontsize{15.5}{23}\selectfont  \bf A Consistency Relation for the Observed Galaxy  Bispectrum }
\end{center}
\begin{center}
{\fontsize{15.5}{23}\selectfont  \bf  and the Local non-Gaussianity  from 
 Relativistic Corrections 
}
\end{center}
\vspace{0.2cm}

\begin{center}
{\fontsize{13}{30} A. Kehagias$^{a,b}$, A. Moradinezhad Dizgah$^{a}$, J. Nore\~na$^{a}$,  H. Perrier$^{a}$ and A. Riotto$^{a}$}
\end{center}

\begin{center}
\vskip 8pt
\textsl{$^{a}$ University of Geneva, Department of Theoretical Physics and Center for Astroparticle Physics (CAP), 24 quai E. Ansermet, CH-1211 Geneva 4, Switzerland }
\vskip 8pt
\textsl{$^{b}$ Physics Division, National Technical University of Athens,15780 Zografou Campus, Athens, Greece }
\end{center}

\vspace{1.24cm}
\hrule \vspace{0.3cm}
{ \noindent \textbf{Abstract} \\[0.2cm] 
\noindent 
}  
We obtain a consistency relation for the observed three-point correlator of galaxies. It includes  relativistic effects and it is valid in the squeezed limit. Furthermore, the consistency relation 
is  non-perturbative and can be  used at arbitrarily small scales for the short modes. Our results are also useful to compute the non-linear relativistic corrections which induce a signal in the observations that might be misinterpreted as primordial  non-Gaussianity with a local shape. 
We estimate the effective local non-Gaussianity parameter from the relativistic corrections. The exact value depends on the redshift and the magnification bias.
At redshift of $z=1.5$, in the absence of magnification bias, we get \hbox {$f^{\rm GR}_{\rm NL} \simeq - 3.0 $}. 
\vspace{0.6cm}

 \hrule

\vspace{0.6cm}

\baselineskip= 16pt

\section{Introduction}  
The experimental developments of large-scale galaxy surveys over the past few decades, in addition to high precision maps of the Cosmic Microwave Background (CMB) have improved extensively our understanding of the universe over a large range of scales. With the upcoming galaxy redshift surveys, such as Euclid \cite{Amendola:2012ys}, the Dark Energy Spectroscopic Instrument (DESI) \cite{Levi:2013gra}, the Large Synoptic Survey Telescope (LSST) \cite{Abell:2009aa} and the Wide-Field InfraRed Survey Telescope (WFIRST) \cite{Spergel:2013tha}, covering progressively a larger fraction of the sky and deeper redshift range, this trend will continue further in the future. 

The enhanced precision of these surveys  requires high accuracy theoretical predictions for the observed quantities rather than theoretically convenient quantities. Moreover,  on the very large-scales probed by these surveys, the Newtonian description of galaxy clustering is not valid. The relativistic description of the observed galaxy number density,  in terms of the observed redshift and angle of galaxies, have been developed up to first order in perturbation theory \cite{Jeong:2011as, Bonvin:2011bg,Challinor:2011bk,Yoo:2010ni,Bruni:2011ta}, which has allowed them to compute the impact of relativistic effects on the galaxy power spectrum. The galaxy bispectrum contains much additional information not present in the power spectrum as there are many more modes and it suffers from different systematics. The computation of this bispectrum in principle needs the use of second order perturbation theory which is very involved and non-trivial. Unlike the CMB,  where the non-linearities are small and therefore (for Gaussian initial conditions) the fluctuations are nearly Gaussian,  the large-scale structure statistics is intrinsically non-Gaussian due to the gravitational collapse. One needs therefore to  go beyond the two-point statistics and hence beyond the linear level in perturbation theory. Recently, several groups have performed the second-order calculation \cite{Yoo:2014sfa, Bertacca:2014dra, Bertacca:2014wga, DiDio:2014lka}. 

The expression for the observed galaxy three-point correlator (bispectrum) obtained from the second-order  calculation is rather complicated and has contributions from various terms. Moreover the region of validity of this calculation is only in the weakly non-linear regime. 

In this paper we point out that a  consistency relation  for the observed galaxy bispectrum can be obtained in
a non-perturbative manner in the squeezed limit (that is one wavenumber much smaller
than the other two in Fourier space). This consistency relation  allows to
express the observed bispectrum in terms of the observed galaxy power spectrum. This is done by correlating the observed galaxy power spectrum in the presence of a long mode with the observed galaxy overdensity on that large wavelength. 
This is exactly computable since the effect of a long mode on the short-scale dynamics is simply a change of frame (or a residual gauge transformation). 

The consistency relations are a consequence of the fact that the  system at hand possesses  non-linearly realized symmetries.  A familiar example of linearly realized symmetry  is the translation invariance of $n$-point correlation functions of,  for example,  the dark matter overdensity field $\delta$.  Non-linearly realized symmetries,  such as  time dependent  spatial transformations,  lead to  more interesting consequences. The first example of a non-trivial consistency relation resulting from a non-linearly realized symmetry was pointed out by Maldacena in the context of single field inflation \cite{Maldacena:2002vr} (see also \cite{Creminelli:2004yq}). In this case the consistency relation is a consequence of the spatial dilation which is non-linearly realized by the curvature perturbations $\zeta$. More recently,   inflationary consistency relations for other non-linearly realized symmetries have been derived \cite{Creminelli:2012ed, Hinterbichler:2012nm, Hinterbichler:2013dpa, Assassi:2012zq, Kehagias:2012pd,Goldberger:2013rsa}. 

In the context of the large-scale structure,  it was first pointed out by Kehagias $\&$ Riotto \cite{Kehagias:2013yd} and Peloso $\&$ Pietroni \cite{Peloso:2013zw} that there are non-linearly realized  symmetries enjoyed by  the system of equations (the continuity, the Euler and the Poisson equations), governing a pressureless fluid coupled to gravity in the Newtonian limit. 
The key point  is that there are large-scale structure observables, {\it e.g.} the dark matter peculiar
velocity, which transform  non-linearly  under some symmetry transformation, while the same transformation shifts the dark matter overdensity  linearly. This  generates consistency relations where the peculiar velocity, and not the overdensity, plays an analogous role to that of the soft-pion in the well-known soft-pion theorem.
The consistency relations are in general  between the squeezed limit $(n+1)$-point function and the $n$-point function of dark matter/galaxy overdensities. In the general relativistic case, these non-linearly realized symmetries that involve the change of the coordinates  can be understood as the result of a residual gauge symmetry. The extension of the consistency relations to the relativistic case has been  done in Refs. \cite{Creminelli:2013mca, Horn:2014rta}.

The advantage of using the  consistency relations in the large-scale structure is that they are directly valid for the observed (and therefore gauge-invariant) quantities, in the same way they are for the  CMB bispectrum \cite{Bartolo:2011wb, Creminelli:2011sq, Kehagias:2014caa, Mirbabayi:2014hda}, and this is what we are going to exploit in the rest of the paper. Our results are useful as a consistency check for the analytical second-order  computations of the bispectrum  of the observed galaxy overdensity and provide a simple and easy way to produce  the squeezed limit. 

Furthermore, since some of these relativistic effects in galaxy number counts give rise to the three point function that mimics that of a local primordial non-Gaussianity (for a review, see Ref. \cite{Bartolo:2004if}), removing their contribution is important for obtaining constraint on primordial non-Gaussianity from galaxy surveys. We will estimate the effective local non-Gaussianity parameter from the non-linear General Relativity (GR) corrections. The exact value depends on the redshift and the magnification bias, see figure \ref{fig:fnl}. At redshift of $z=1.5$ and in the absence of the magnification bias we get \hbox {$f^{\rm GR}_{\rm NL} \simeq - 3.0 $}. 

This paper is organized as follows. In section \ref{sec:coord} we review the coordinate transformation that induces the long wavelength mode and continue in section \ref{sec:gal_long} to describe the effect of the long wavelength perturbation mode on the observed galaxy overdensity. We use this result to calculate the bispectrum in the squeezed limit by correlating two short-mode overdensities with a long wavelength mode. We then decompose the bispectrum in terms of spherical harmonics in section \ref{sec:SH}.  In section \ref{sec:fnl} we estimate the value of the effective $f^{\rm GR}_{\rm NL} $ induced by these relativistic corrections. Finally,  we conclude in section \ref{sec:conc}.

\section{Adiabatic modes and residual gauge symmetry} \label{sec:coord}

The squeezed limit of an $(n+1)$-point correlation function in which one of the scales is much larger than the other $n$ can be written as the correlation of a long-wavelength mode with the $n$-point function in the presence of that long mode, {\it e.g.}
\be
\Big< \Phi(\lambda_L) \Delta_{\rm g}(\lambda_1)\dots\Delta_{\rm g}(\lambda_n)\Big>_{\lambda_L \gg \lambda_i}= \Big<\Phi(\lambda_L) \Big<\Delta_{\rm g}(\lambda_1) \dots \Delta_{\rm g}(\lambda_n)\Big>_L\Big>\, ,
\ee
where $\Phi$ is Bardeen's potential, $\Delta_{\rm g}$ is the galaxy number overdensity, $\lambda$ is the characteristic scale of each variable, and $\langle\cdots\rangle_L$ is to be interpreted in the conditional probability sense: it is an average given the condition that when smoothed at large-scales Bardeen's potential takes the value $\Phi(\lambda_L)$.

It thus suffices to compute the effect of the long-wavelength potential on the short scale dynamics. This task is greatly simplified by the fact that on large enough scales,  one can approximate Bardeen's potential as a Taylor expansion
\be
\Phi_L(\tau, \vec{x}) = {\Phi}_L(\tau,\vec{x}_{\cal E}) + (\vec{x}-\vec{x}_{\cal E})\cdot\vec{\nabla}{\Phi}_L(\tau,\vec{x}_{\cal E})\, ,
\label{philong}
\ee
where we have centered the Taylor expansion around an arbitrary point close to where we evaluate the short-scale $n$-point function. 
We will now show that in a $\Lambda$CDM universe this large-scale potential is a gauge mode: an appropriate coordinate transformation can set these first two terms of the Taylor expansion to zero. We will summarize the derivation given in Refs. \cite{Creminelli:2012ed, Creminelli:2013mca}. 

We will work to first order in the gradient of the long mode, which means we will ignore its curvature $\nabla^2\Phi_L$ and tidal forces $\de_i\de_j\Phi_L$. Though these effects are in general large at sub-horizon scales, their effect would not be confused in observations with a violation of the consistency relation connecting the squeezed limit of the 3-point function with the power spectrum. For simplicity, we also choose to be perturbative in the short-mode potential $\Phi_{\cal E}$ and velocities $\sim\de_i\Phi_{\cal E}$ which are expected to be small even at very small scales. Note that we never assume the short-mode density $\sim \nabla^2\Phi_{\cal E}$ to be small and in this sense we are non-perturbative.

Let us start by writing the perturbed FLRW metric in Poisson gauge
\be
\d s^2 = a^2(\tau)\Big[-(1+2\Phi)\d\tau^2 + (1 - 2\Psi){\rm d}\vec{x}^2\Big]\,,
\label{eq:perturbedFLRW}
\ee
where $\Phi$ and $\Psi$ are Bardeen's potentials and the Poisson gauge condition fixes the time-space part of the metric to be zero and the tensor part (which we ignore for simplicity) to satisfy $\partial_i \partial_j \gamma^{ij} = 0$. Now,  consider the following coordinate transformation
\begin{align}
\label{restau}
\tau &\mapsto \widetilde\tau = \tau + \epsilon(\tau) + \vec{x}\cdot\vec{\xi}'(\tau) + \alpha(\tau, \vec{x})\,, \\
\label{resx}
x^i &\mapsto \widetilde{x}^i = x^i(1 + \lambda + 2\vec{x}\cdot\vec{b}) - b^i \vec{x}^2 + \xi^i(\tau)\, 
\end{align}
where $\alpha$ satisfies $\vec{\nabla}^2 \alpha = -2\partial_i(\Phi + \Psi)\xi'^i$, $\lambda$ and $\vec{b}$ are constants, and $\epsilon$ and $\vec{\xi}$ are functions of time. 

The key point is that,  after performing this transformation, the metric is still of the form given by Eq. \eqref{eq:perturbedFLRW}.
The potentials are now
\begin{align}
\widetilde\Phi &= \Phi + \big[\epsilon' + \vec{x}\cdot\vec{\xi}'' + \mathcal{H}(\epsilon + \vec{x}\cdot\vec{\xi}')\big]\,,\label{eq:phi}\\
\widetilde\Psi &= \Psi  - \big[\lambda + 2\vec{x}\cdot\vec{b} + \mathcal{H}(\epsilon + \vec{x}\cdot\vec{\xi}')\big]\,.\label{eq:psi}
\end{align}
Note that the transformation leaves the dynamics unchanged, while it changes perturbation quantities. Therefore, since these perturbations are still described by the same gauge condition, the symmetry at issue is a non-linearly realized symmetry. The transformation we have performed  is a residual gauge symmetry of the Poisson gauge\footnote{A residual gauge symmetry  is of course there for  any other gauge different from the Poisson one, which we have chosen because it is usually adopted as a starting point in the
computation for the large-scale structure.}.
This means that even when one completely fixes the gauge to be the Poisson gauge in which the quantities appearing in the metric are Bardeen's potentials, there is a residual gauge freedom under some coordinate transformations that do not leave the background invariant, such as Eqs. (\ref{restau}, \ref{resx}). 
From now on will ignore $\alpha$ in Eq. \eqref{restau} since its effect is of higher order under our approximations.

An arbitrary coordinate transformation that does not leave the background FLRW metric invariant can be interpreted as inducing perturbations of the metric which would have no physical meaning whatsoever. However, in a context where the curvature perturbation $\zeta$ is conserved, such as for $\Lambda$CDM even deep inside the horizon, the transformations (\ref{restau}, \ref{resx}) can be chosen to solve the large wavelength limit of Einstein's equations.  
In order to satisfy this, the transformation parameters obey the following conditions \cite{Creminelli:2012ed, Creminelli:2013mca},
\be 
v_L = -(\epsilon + \vec{x}\cdot\vec{\xi}')\,,
\label{eq:v}
\ee
where $v$ is the velocity potential $v^i = \partial_i v$. We see that $\vec{\xi}'$ corresponds to a long-wavelength velocity mode. Additionally we require,
\be
\epsilon' + 2\mathcal{H}\epsilon = -\lambda\,,\qquad\vec{\xi}'' + 2\mathcal{H}\vec{\xi}' = -2 \vec{b}\,.
\ee
In order to see when these conditions are consistent, we note that from the definition of the comoving curvature perturbation $\zeta \equiv -\Psi + \mathcal{H}v$ and from equation \eqref{eq:psi} we have
\be
\zeta_L = \lambda + 2\vec{x}\cdot\vec{b}\,.
\label{eq:zeta}
\ee
Thus, \emph{a long-wavelength solution is a gauge mode if on those large-scales the comoving curvature perturbation is constant in time}. This means that the long-wavelength limit of a physical solution of Bardeen's potentials is equivalent to a coordinate transformation and it is what Weinberg calls adiabatic modes \cite{Weinberg:2003sw}. Indeed, since $\zeta$ is conserved outside the sound horizon which is nearly zero in a $\Lambda$CDM universe, this would hold up to scales where baryonic physics start being important. Note that this constancy of $\zeta$ must be satisfied only by the long-wavelength mode for our arguments to be valid; the short-wavelength modes can be at any scale.

\section{Observed galaxy overdensity in the presence of long mode } \label{sec:gal_long}

In this section we compute the effect of a large-scale mode approximated by a constant plus a gradient as in Eq. \eqref{philong}. We start by defining the galaxy number density, we then include the effect of the long mode and we show that if the long mode has a characteristic scale that is much larger than all other scales in the problem (including the distance between the observer and the galaxies), its effect is zero as expected.

\subsection{Galaxy number density}

In galaxy surveys, the observed galaxy number density at a given redshift and angle on the sky $n^{\mathrm{obs}}_{\rm g}$, is obtained by counting the number of galaxies, ${\rm d}N_{\rm g}$  within the observed volume, ${\rm d}V = {\rm d}\Omega {\rm d}z$ 
\be
n_{\rm g}^{\mathrm{obs}}(z,\n) = \frac{{\rm d}N_{\rm g}(z,\n)}{{\rm d}V}. 
\ee 
From here, one defines the galaxy number overdensity $\Delta_{\rm g}$ as
\be
n_{\rm g}^{\mathrm{obs}}(z,\n) \equiv \bar{n}_{\rm g}(z)(1 + \Delta_{\rm g}(z,\n))\,.
\ee
where $\bar{n}_{\rm g}(z) \equiv \langle n_{\rm g}^{\mathrm{obs}}(z)\rangle$ is the observed number density of galaxies averaged over the angle at a fixed redshift.

In order to study the effect of a long-wavelength mode on the galaxy number density, we must first relate it to the physical density. In the inhomogenous universe, the observed redshift and the propagation direction of the photons differ from the true redshift and angle. Therefore the observed volume defined in terms of the observed redshift and solid angle differs from the physical volume occupied by the source 
\be
dV_{\rm phys} = {\rm d}V_{\rm obs} (1+ \delta V)
\ee
Thus the observed number density of galaxies which is obtained by counting the number of galaxies within the observed volume differs from the physical number density by a factor due to the volume distortions 
\be
n_{\rm g}^{\mathrm{obs}}(z,\hat{n}) = n_{\rm g}^{\mathrm{phys}}(z,\hat{n})(1+\delta V
)\,.
\label{physobs}
\ee
Moreover, galaxy surveys are flux limited, {\it i.e.}, only galaxies above some threshold luminosity are observed. This threshold luminosity is inferred from the observed flux 
\be
\hat L = 4\pi  \bar {\mathcal{D}}_l^2(z) f_{\rm obs} ,
\ee
where the luminosity distance $\bar{\mathcal{D}}_l(z) $ is that in a homogenous universe and $\hat L$ is the inferred luminosity. However the physical luminosity of the galaxy is different than the inferred luminosity since the luminosity distance to the galaxy is not that of the homogenous universe and the photon propagation is affected by the fluctuations along the path 
\be
L^{\rm phys} = 4 \pi \mathcal{D}_l(z)^2 f_{\rm obs} = 4 \pi \bar {\mathcal{D}}_l(z)^2 f_{\rm obs}  (1+\delta \mathcal{D}_l)^2 = \hat L(1+\delta \mathcal{D}_l)^2
\ee
Therefore the observed number of galaxies at observed redshift $z$, angle $\hat n$ and above a threshold $L$ is related to the physical number count as
\begin{align}\label{eq:n_obs}
n_{\rm g}^{\mathrm{obs}}(z,\hat{n}, \hat L) &= n_{\rm g}^{\mathrm{phys}}(z,\hat{n}, \hat L(1+\delta \mathcal{D}_l)^2)(1+\delta V)\, \nonumber \\
&= n_{\rm g}^{\rm phys}(z,\hat n)(1+\delta V)(1+t\delta \mathcal{D}_l)
\end{align}
where $t= 2 \frac{d \ln \bar n^{\rm phys}_{\rm g}(z)}{d \ln L}$. In writing the second line in Eq. (\ref{eq:n_obs}), as explained in the beginning of this section, we made two assumptions: first we only kept the linear contributions to the luminosity distance and volume perturbations. This is because if we neglect the strong lensing, the perturbations in the luminosity distance and volume are proportional to the gravitational potential and its gradient and thus their higher order contributions are subdominant with respect to the higher order contribution of the matter density. Thus we neglect these terms. Second in writing the contribution from the magnification bias, we assumed that the perturbations and the background have the same dependance on the luminosity.  

\subsection{Long-short couplings}

We now wish to compute the galaxy over-density measured in a small patch of the sky and in a small range of redshifts, in the presence of a mode of Bardeen's potential which is much larger than the size of this patch. Following the arguments of section \ref{sec:coord}, this mode should have no effect on the local physics of that small region since it is equivalent to a coordinate transformation. However, we take here this large-scale mode to have a characteristic size which is smaller than the distance between the observer and the small patch, which means that it will indeed have an effect on the correlations measured by the observer. This effect is purely due to the fact that we measure the galaxy correlation functions in terms of the redshift and angular position of the galaxies rather than the physical separation between galaxies and their proper time. For this reason this non-zero correlation is a ``projection effect''.

In the presence of the long mode, the observed redshift and angle of the galaxies are modified with respect to those in the absence of long mode.  First lets compute the effect of the large-scale mode on the observed angular position of the galaxy 
\be
\n_{\cal O}^i \equiv \frac{-T_\mu^iP^\mu(\tau_{\cal O})}{\sqrt{\delta_{jk} T_\nu^jP^\nu T_\eta^k P^\eta}}\,,
\ee
where $T_\mu^i P^\mu = (\delta^i_\mu + u^{{\cal O}\,i} u^{\cal O}_{\phantom{o}\,\mu})P^\mu$ is the projection of the photon 4-momentum $P^\mu$, orthogonal to the observer's 4-velocity $u^{\cal O}$. At linear order the solution to the photon geodesic gives
\be
\hat{n}_{\cal O} = \hat{n}_{\cal E} + \vec{v}^\perp_{{\cal O}} - \int_{\tau_{\cal O}}^{\tau_{\cal E}}\mathrm{d}\tau'\,\vec{\nabla}^{\perp}\left(\Phi(\tau') + \Psi(\tau')\right)\,,
\label{no}
\ee
where the direction of emission is defined as $\n_{\cal E} \equiv -P^i(\tau_{\cal E})/\sqrt{\delta_{jk}P^j(\tau_{\cal E})P^k(\tau_{\cal E})}$ and the superscript ${}^\perp$ denotes a projection orthogonal to $\n_{\cal E}$, e.g. $v^{i\,\perp} = (\delta^{ij} - \n_{\cal E}^i \n_{\cal E}^j) v^j$. The second term on the right hand side corresponds to an aberration due to the motion of the observer and the last term is the lensing of the photon trajectory.

The presence of the large-scale mode will contribute to the lensing integral in Eq. \eqref{no}, and it will also cause the photon to be emitted in a different direction (in the coordinate system with the long mode). This change in the direction of emission can be computed simply by remembering that at the point of emission the large-scale mode is equivalent to a coordinate transformation which will transform the vector $P^i(\tau_{\cal E})$ giving
\be
\n_{\cal E} \mapsto \hat{\widetilde{n}}_{\cal E} = \hat{n}_{\cal E} - \vec{\xi}'^\perp(\tau_{\cal E}) + 2 \vec{b}^\perp(\tau_{\cal E} - \tau_{\cal O})\,,
\ee
where we used the fact that at zeroth order $\vec{x}_{\cal E} = -\n_{\cal E}(\tau_{\cal E} - \tau_{\cal O})$. Writing $\vec{\xi}$ and $\vec{b}$ in terms of the potentials as in Eqs. (\ref{eq:v}) and (\ref{eq:zeta}) and isolating the contribution of the long mode to the lensing integral in Eq. \eqref{no} we get
\be
\hat{\widetilde n}_{\cal O} = \hat{n}_{\cal O} -\vec{v}^\perp_{\cal O} + (1 + \mathcal{H}(\tau_{\cal E} - \tau_{\cal O}))\vec{v}^\perp_L(x_{\cal E})  - (\tau_{\cal E} - \tau_{\cal O})\nabla^\perp\Psi_L(x_{\cal E}) + \int_{\tau_{\cal O}}^{\tau_{\cal E}}\mathrm{d}\tau'\,\nabla^{\perp}\left(\Phi_L(\tau') + \Psi_L(\tau')\right)\,.
\label{deltan}
\ee
The redshift is defined as the ratio of the energies of a photon at emission and observation 
$$(1+z) \equiv \frac{P^\mu(\tau_{\cal E})u_{{\cal E}\,\mu}}{P^\nu(\tau_{\cal O})u_{{\cal O}\,\nu}}\,,$$
and a calculation analogous to the one performed for the direction gives
\be
\widetilde z = z  +(1+ z)\bigg[\mathcal{H}v_L(x_{\cal E}) - \mathcal H_{\cal O} v_{\cal O}+ \Phi_L(x_{\cal E}) -\Phi_{\cal O} - \n\cdot(\vec{v}_L(x_{\cal E})-\vec{v}_{\cal O}) - \int_{\tau_{\cal O}}^{\tau_{\cal E}} \mathrm{d}\tau'\,(\Phi_L'+\Psi_L')(\tau')\bigg]\,.
\label{deltaz}
\ee
The observed number density of galaxies in the presence of the long mode is therefore given by
\begin{align}\label{eq:n}
\left.n_{\rm g}^{\mathrm{obs}}(z,\hat{n}, \hat L)\right |_L &= n_{\rm g}^{\mathrm{phys}}(z+\Delta z,\hat{n}+\Delta \hat n) (1+\delta V^{(L)})(1+\delta V^{(S)})(1+t\delta \mathcal{D}_l^{(S)})(1+t\delta \mathcal{D}_l^{(L)}) \nonumber \\
&=  \bar n_{\rm g}(z)\left[1+\Delta_{\rm g}(z+\Delta z,\hat{n}+\Delta \n)\right]\left(1+ e \Delta z + \delta V^{(L)} +  t \delta \mathcal{D}_l^{(L)}\right),
\end{align}
where\footnote{In the literature the evolution bias is often defined as $ f_{\rm evo} \equiv \frac{\d\ln(a^3 \bar{n}^{\mathrm{phys}}_{\rm g})}{\mathcal{H}\d\tau_{\cal E}} $, therefore $f_{\rm evo} = 3 - e/a$.} $e\equiv\frac{{\rm d} \ln \bar n^{\rm phys}_{\rm g}}{{\rm d}z}$, and $\Delta_{\rm g}$ has contributions from the luminosity distance and volume perturbations due to the short modes as well as the density perturbations due to the short modes.  Note that $\Delta z = \widetilde z - z$ and $\Delta \n = \hat{\widetilde n} - \n$ are the corrections to the observed angle and redshift due to the long mode. Expanding Eq. (\ref{eq:n}) linearly the galaxy number overdensity is finally given by
\vspace{0.4cm}
\begin{mdframed}
\begin{align}
\Delta_{\rm g}(z, \n)|_L &= \Delta_{\rm g}( z, \hat{n}) + (1 + \Delta_{\rm g}(z, \hat{n}))\left(e \Delta z+t\delta \mathcal D_l^{(L)}+ \delta V^{(L)}\right) \nonumber  \\
&+ \Delta z \frac{\de}{\de z}\Delta_{\rm g}(z, \hat{n}) + \Delta\n \cdot \frac{\de}{\de\n} \Delta_{\rm g}(z, \hat{n}) + \mathcal{O}\bigg(\frac{\Phi_L}{\lambda_L^2 \mathcal{H}^2}\Delta_{\rm g}\bigg)\,,\nonumber
\end{align}
\end{mdframed}
\be
\label{eq:deltagL}
\ee
For our purposes it suffices to use the perturbative expressions for the quantities $\delta V^{(L)}$ and $\delta \mathcal{D}_l^{(L)}$ at first order \cite{Yoo:2009au, Bonvin:2011bg, Jeong:2011as}
\begin{eqnarray}
\delta V^{(L)} = & -&2 (\Phi_L+ \Psi_L )  + \frac{1}{\mathcal{H}} \Psi_L' + \left( \frac{\mathcal{H}'}{\mathcal{H}^2}+\frac{2}{r \mathcal{H}}\right)	\Phi_L\nonumber \\
& + & \left(-3+\frac{\mathcal{H}'}{\mathcal{H}^2} +\frac{2}{r \mathcal{H}}\right) \left(	-\vec{v}_L\cdot \hat{n}  + \int_\tau^{\tau_{\cal  O}} \d\tau' (	\Phi_L' + \Psi_L' ) (\tau') \right) \nonumber  \\
&  +& 	\int_\tau^{\tau_{\cal  O}}\d\tau'  \left(\frac{2}{r}-\frac{r-r(\tau')}{rr(\tau')}\Delta_\Omega \right) (\Phi_L + \Psi_L)(\tau')\,,
\end{eqnarray}
and
\begin{eqnarray}
\delta \mathcal{D}_l^{(L)} &=& \left(\frac{1}{r \mathcal{H}}-1 \right) \left(-\vec{v}_L \cdot \hat{n} + \Phi_L + \int_\tau^{\tau_{\mathcal{O}}} \d\tau'\, (\Psi_L' + \Phi_L')(\tau') \right) \nonumber \\
&+& \frac{1}{2}\int_\tau^{\tau_{\mathcal{O}}} \d \tau' \left[ \frac{2}{r}- \frac{r-r(\tau')}{r r(\tau')}\Delta_\Omega \right] \left(\Psi_L + \Phi_L \right)(\tau')  - \Phi_L\,,
\end{eqnarray}
where all the quantities without an argument are evaluated at the point of emission.

The last step we need to take is to correlate the expression (\ref{eq:deltagL}) with another short-scale $\Delta_{\rm g}$ and a long-scale one, and go to harmonic space to obtain the consistency relation of the galaxy three-point function in the squeezed limit
\begin{eqnarray}
\Big< \Delta^L_{\rm g}(z_1, \n_1) \Deltag{2} \Deltag{3} \Big> &= & \bigg[\Big< \Delta_{\rm g}^L(z_1, \n_1) 
d(z_2, \n_2 )\Big> 
+ \Big< \Delta_{\rm g}^L(z_1, \n_1) \Delta z(z_2, \n_2)\Big>\frac{\de}{\de z_2} \nonumber\\
&+& \Big< \Delta_{\rm g}^L(z_1, \n_1) \Delta\n(z_2, \n_2)\Big>\cdot\frac{\de}{\de\n_2}\bigg]\Big< \Delta_{\rm g}(z_2, \n_2) \Delta_{\rm g}(z_3, \n_3)\Big>\nonumber\\
&+& (2 \leftrightarrow 3)\,,
\label{eq:master}
\end{eqnarray}
where we have introduced 
\begin{equation}
d \equiv (e \Delta z+t\delta \mathcal D_l^{(L)}+ \delta V^{(L)}).
\end{equation}

Let us close this section by commenting on the corrections to Eq. \eqref{eq:deltagL}. In general, we also expect corrections going like the curvature induced by the long-wavelength mode, that is $(\Phi_L \Delta_{\rm g}/ \lambda_L^2\mathcal{H}^2)$.  Parametrically, these are subdominant with respect to the lensing and redshift space distortions which behave as $(\Phi_L \Delta_{\rm g}/\lambda_L \lambda_S {\mathcal H}^2)$, but they are larger than the other relativistic corrections. However as we will discuss in section \ref{sec:second-order}, there are additional suppressions for the lensing and the redshift space distortions. In spite of this, our result is a consistency relation in the sense that if the evolution of the universe is adiabatic (single-field inflation and a subsequent evolution that conserves $\zeta$ on the scales of interest) our relation captures the behavior of the squeezed limit as we take $1/\lambda_L$ to zero up to first order in $\lambda_S/\lambda_L$. Indeed, the terms we computed could be mistaken for a non-zero local non-Gaussianity, which corresponds to a physical coupling between $\Phi_L$ and $\Phi_{\cal E}$, or a signal of violation of the Equivalence Principle if they are not taken into account. Furthermore, there are several approaches to compute these curvature terms in the literature, which correspond to computing the evolution of the short modes in a curved universe \cite{Baldauf:2011bh,Creminelli:2013cga, Dai:2015rda} (see also \cite{Kehagias:2013paa,Valageas:2013zda}) but they can only be approximate since it is in general impossible to analytically compute the effect of curvature on the evolution of these non-perturbative short modes. 

Note that in real observations, one counts the number of galaxies within a finite redshift bin which have two effects. First, the characteristic scale of the short mode is determined both by its angular scale and the assumed width of the redshift slices. Therefore the curvature terms are only negligible if $\lambda_L \gg \delta z/ {\mathcal H}$.  Second as we will explain in more details in section \ref{sec:SH}, in order to relate the theoretically calculated bispectrum to the one measured in a survey, one needs to integrate over the window functions describing the redshift bins. Note that the consistency relation \emph{does not break down} when performing this integration as long as the characteristic spatial scale of the redshift bin is smaller than the characteristic scale of the long-wavelength mode. This is easily achieved especially for large redshifts $z \gtrsim 1$. 

\subsection{Validity checks}

Let us now perform various validity checks on our results.

\begin{itemize}

\item We expect a large mode to have no observable effect when its scale is larger than all the scales in the problem, \emph{i.e.} the distance between the source galaxies and the distance between them and the observer. This is similar to what happens for the CMB \cite{Creminelli:2011sq,Mirbabayi:2014hda}. One way to check that our calculation is consistent with this fact is to perform a coordinate transformation that induces such a large-scale mode and check that the galaxy number over-density remains invariant. Let us begin by checking that this is indeed the case for $\hat n_{\cal O}$, setting the observer's position to zero $\vec{x}_{\cal O} = \vec{0}$, we get
\be
\hat{\widetilde n}_{\cal O} = \n_{\cal O} + \vec{\xi}'^\perp(\tau_{\cal O}) - \vec{\xi}'^\perp(\tau_{\cal E}) + 2(\tau_{\cal E} - \tau_{\cal O})\vec{b}^\perp + \int^{\tau_{\cal E}}_{\tau_{\cal O}}\d\tau'\left(\vec{\xi}''(\tau') - 2\vec{b}\right)^\perp\,,
\ee
where all the terms of the transformation cancel. An analogous argument holds for the redshift $z$, and we get
\begin{eqnarray}
\widetilde z &=& z +(1+z)\bigg[\epsilon'(\tau_{\cal E}) + \vec{x}_{\cal E}\cdot\vec{\xi}''(\tau_{\cal E}) + \n\cdot\vec{\xi}'(\tau_{\cal E}) -\epsilon'(\tau_{\cal O}) - \n\cdot\vec{\xi}'(\tau_{\cal O}) \nonumber\\
&-& \int_{\tau_{\cal O}}^{\tau_{\cal E}}\d\tau'\,\left(\epsilon'(\tau')+\vec{x}\cdot\vec{\xi}''(\tau')\right)'\bigg]\,,
\end{eqnarray}
where all the terms in the parenthesis cancel after writing the partial time derivative in terms of a total derivative in the integral $\de_\tau = \d/\d\tau + \n\cdot\vec{\nabla}$. For these cancellations to happen it is important to replace the Bardeen's potentials $\Phi$ and $\Psi$ inside the integral by an exact constant plus a gradient, Eq. \eqref{philong}, which does not hold if the long mode oscillates between the observer and the source, \emph{i.e.} if the characteristic scale of the long mode is smaller than the distance between the observer and the source. We thus obtain
\be
n_{\mathrm{g}}^{\mathrm{obs}}(z,\n)|_L = \bar{n}_{\mathrm{g}}(z)\left(1+\Delta_{\mathrm{g}}(z,\n)\right)\left(1+\delta V^{(L)} + 2t\delta\mathcal{D}_l^{(L)}\right)\,.
\ee
Finally, the average in this equation $\bar{n}_{\mathrm{g}}$ assumes that the observer has access to the large-scale mode, such that the terms in the second parenthesis give zero when averaged over angles. However, when the mode is much larger than all the scales in the problem, the angular averages of these terms do not vanish,
\be
\langle n^{\mathrm{obs}}_{\mathrm{g}}(z,\n)\rangle|_L = \langle \bar{n}_{\mathrm{g}}(z)\rangle \left(1+\langle \delta V^{(L)}\rangle + \langle 2t\delta\mathcal{D}_l^{(L)}\rangle\right)\,,
\ee
but they are zero-modes on the observed patch and they can be reabsorbed in the average of the galaxy number density
\be
n^{\mathrm{obs}}_{\mathrm{g}}(z,\n)|_L =\langle n^{\mathrm{obs}}_{\mathrm{g}}(z,\n)\rangle|_L\left(1+\Delta_{\mathrm{g}}(z,\n)\right)\,.
\ee
We thus conclude that a mode with a characteristic scale much larger than all the scales in the problem has no effect on the $\Delta_{\rm g}$. We remark once more that \emph{when the large-scale mode is smaller than the distance between the source and the observer, these contributions will all be non-zero}.

\item As a further check of the validity of our calculation we take the super-horizon limit in equation \eqref{eq:deltagL} and work to first order in the long modes and zeroth order in the short modes. We expect to obtain the same results as the linear calculation of Ref. \cite{Jeong:2011as, Bonvin:2011bg,Challinor:2011bk,Yoo:2010ni}, which in a pressureless medium is given by
\be
\Delta_{\rm g}^{(1)} = b \delta_{\rm m, sync}+ \left(-e \delta z_{\rm sync}+t\delta \mathcal D_L+ \delta V\right)\, \nonumber ,
\ee  
 where $\delta_{\rm m,  sync}$ and $\delta z_{\rm sync}$ are the matter over density and redshift perturbations in the synchronous gauge. Indeed, $\Delta z = -\delta z_{\rm sync}$ and at super-horizon scales $\delta_{\rm m, sync}$ is suppressed by $1/\lambda_L^2\mathcal{H}^2$, and we recover our result.
 
 \item As remarked in the literature, the expressions for $\delta V^{(L)}$ and $\delta \mathcal{D}_l^{(L)}$ are gauge invariant. The reader might worry that $\Delta z$ and $\Delta \n$ are not since they explicitly involve the velocity which is not gauge invariant. However, it is straightforward to write them at first order in terms of explicitly gauge invariant variables
 \begin{multline}
 \Delta\n = \left(\frac{1}{\mathcal{H}(\tau_{\cal E})} + (\tau_{\cal E} - \tau_{\cal O})\right)\vec{ \nabla}^\perp\zeta_L(x_{\cal E})+\frac{1}{\mathcal{H}(\tau_{\cal E})}\vec{\nabla}^\perp\Psi_L(x_{\cal E}) \\+ \int_{\tau_{\cal O}}^{\tau_{\cal E}}\mathrm{d}\tau'\,\vec{\nabla}^{\perp}\left(\Phi_L(\tau') + \Psi_L(\tau')\right)\,,
 \end{multline}
 \be
\Delta z = (1+ z)\bigg[\zeta_L(x_{\cal E}) + \Psi_L(x_{\cal E}) + \Phi_L(x_{\cal E}) - \frac{1}{\mathcal{H}}\n\cdot\nabla(\zeta_L + \Psi_L)(x_{\cal E}) - \int_{\tau_{\cal O}}^{\tau_{\cal E}} \mathrm{d}\tau'\,(\Phi_L'(\tau')+\Psi_L'(\tau'))\bigg]\,,
 \ee
 where we have dropped velocities and potentials evaluated at the observer as we will do in the rest of this paper.

\end{itemize}

\section{Spherical Harmonic Decomposition}\label{sec:SH}
In both N-body simulations and data from galaxy surveys, the three-point function is often measured and analyzed in Fourier-space. This is a convenient decomposition if the galaxies have similar redshifts. However here we are interested in large-scales for the long mode where the relativistic effects can be important. Hence a single Fourier mode of large wavelength may include galaxies with significantly different redshifts. Due to the evolution of the universe, there is no translation invariance in the radial direction of the redshift space, {\it i.e.} $z$. Therefore the Fourier space bispectrum is not proportional to the Dirac delta:  triangles do not close. We thus choose to decompose the three-point function in spherical harmonics, which could prove advantageous or even necessary for this type of analysis, 
\begin{eqnarray}
\Big< \alm{1} \alm{2} \alm{3} \Big>_{\ell_1 \ll \ell_2, \ell_3}&= &\int \mathrm{d}\Omega_1 \mathrm{d}\Omega_2 \mathrm{d}\Omega_3\, \Ys{1} \Ys{2} \Ys{3} \nonumber \\
&\times& \Big< \Delta^L_{\rm g}(z_1 , \n_1) \Deltag{2} \Deltag{3} \Big> \, .
\end{eqnarray}
This basis naturally fits the sphere and does not have the problem of triangles not closing. We now compute each of the terms contributing to the above equation in turn. The contribution of the terms in the first line of Eq. \eqref{eq:master} are easy to compute and give
\begin{eqnarray}
\Big< \alm{1} \alm{2} \alm{3} \Big>_{\ell_1 \ll \ell_2, \ell_3}&\supset& \gaunt \bigg[C_{\ell_1}^{\Delta_{\rm g} d}(z_1, z_2) + C_{\ell_1}^{\Delta_{\rm g} \Delta z}(z_1, z_2)\frac{\de}{\de z_2}\bigg]\nonumber\\ 
&\times & C_{\ell_3}^{\Delta_{\rm g} \Delta_{\rm g}}(z_2, z_3) + (2 \leftrightarrow 3) \ ,
\end{eqnarray}
where the Gaunt integral is defined as
\begin{eqnarray}
\gaunt &=&  \int \mathrm{d}\Omega  \ Y_{\ell_1 m_1}(\n)\ Y_{\ell_2 m_2}(\n)\ Y_{\ell_3 m_3}(\n) \nonumber \\
&=& \begin{pmatrix}\ell_1 & \ell_2 & \ell_3\\ m_1 & m_2 & m_3\end{pmatrix}
\begin{pmatrix}\ell_1 & \ell_2 & \ell_3\\ 0 & 0 & 0\end{pmatrix}\sqrt{\frac{(2\ell_1 + 1)(2\ell_2 + 1)(2\ell_3 + 1)}{4\pi}}.
\end{eqnarray}
The third line of Eq.  \eqref{eq:master} is slightly more complex
\begin{multline}
\Big< \alm{1} \alm{2} \alm{3}\Big>_{\ell_1 \ll \ell_2, \ell_3}\supset \int \mathrm{d}\Omega_1 \mathrm{d}\Omega_2 \mathrm{d}\Omega_3\, \Ys{1} \Ys{2} \Ys{3}\\
\times \frac{\de}{\de \n_2} \Big< \Delta_{\rm g}(z_1, \n_1)I(z_2,\n_2) \Big>\cdot\frac{\de}{\de\n_2}\Big< \Delta_{\rm g}(z_2, \n_2) \Delta_{\rm g}(z_3, \n_3)\Big> +(2\leftrightarrow 3)
\end{multline}
where
\be
I(z_2,\n_2) \equiv \Psi_L +\left(\frac{1}{r(\tau_2)} - \mathcal{H}(\tau_2)\right)v_L+\int_{\tau_{\cal O}}^{\tau_2}\frac{\d\tau'}{r(\tau')}(\Phi_L + \Psi_L) (\tau', \n_2r(\tau')).
\ee
This can be further simplified by decomposing the two-point functions in spherical harmonics, giving
\begin{multline}
\Big< \alm{1} \alm{2} \alm{3} \Big>_{\ell_1 \ll \ell_2, \ell_3}\supset \int \mathrm{d}\Omega\, \bigg(\frac{\de}{\de \n^i}\ys{1}\bigg) \ys{2} \bigg(\frac{\de}{\de \n^i}\ys{3}\bigg) \\
 \times C^{\Delta_{\rm g} I}_{\ell_1}(z_1, z_2) C^{\Delta_{\rm g} \Delta_{\rm g}}_{\ell_3}(z_2, z_3)
+ (2\leftrightarrow 3)\,.
\end{multline}
The integral in the first line can be rewritten as
\begin{equation}
\frac{1}{2}\int \mathrm{d}\Omega \Big[\Delta_\Omega\big(\ys{1}\ys{3}\big) - \big(\Delta_\Omega\ys{1}\big)\ys{3} - \ys{1}\big(\Delta_\Omega\ys{3}\big)\bigg]\ys{2}\,.
\end{equation}
The second and third terms in the parenthesis give simply
\begin{equation}
\frac{1}{2}\gaunt(\ell_1(\ell_1 + 1) + \ell_3(\ell_3 + 1)).
\end{equation}
We write the product of spherical harmonics of the first term as a Clebsch-Gordan decomposition
\begin{eqnarray}
&& \frac{1}{2}\int \mathrm{d}\Omega \bigg[\Delta_\Omega\sum_{LM}\sqrt{\frac{(2\ell_1 + 1)(2\ell_3 + 1)(2L + 1)}{4\pi}}
\begin{pmatrix}\ell_1 & \ell_3 & L\\ m_1 & m_3 & M\end{pmatrix}
\begin{pmatrix}\ell_1 & \ell_3 & L\\ 0 & 0 & 0\end{pmatrix}
Y_{LM}(\n)\bigg]\ys{2} \nonumber \\
&& = -\frac{1}{2}\ell_2(\ell_2 + 1)\gaunt\,.
\end{eqnarray}
Putting everything together for the observed reduced bispectrum of galaxies, $b_{\ell_1,\ell_2,\ell_3}(z_1,z_2,z_3)$, defined as 
\be
\Big< \alm{1} \alm{2} \alm{3} \Big> = \gaunt \ b_{\ell_1,\ell_2,\ell_3}(z_1,z_2,z_3),
\ee
we obtain the main result of this paper, namely the consistency relation in the squeezed limit in multipole space 
\vskip 0.4cm
\begin{mdframed}
\begin{multline}
\lim_{\ell_1\ll \ell_2,\ell_3}  b_{\ell_1,\ell_2,\ell_3}(z_1,z_2,z_3) =  \bigg[C_{\ell_1}^{\Delta_{\rm g} d}(z_1, z_2) + C_{\ell_1}^{\Delta_{\rm g} \Delta z}(z_1, z_2)\frac{\de}{\de z_2}  \\
+\frac{1}{2}\left( \ell_1(\ell_1 + 1)  - \ell_2(\ell_2 + 1) 
+\ell_3(\ell_3 + 1)\right)C^{\Delta_{\rm g} I}_{\ell_1}(z_1, z_2) \bigg] C^{\Delta_{\rm g} \Delta_{\rm g}}_{\ell_3}(z_2, z_3)  + (2 \leftrightarrow 3). \nonumber
\end{multline}
\end{mdframed}
\be\label{final}
\ee

As pointed out before, since in real observation, one counts the number of galaxies within a finite redshift bin, in order to compare with observed galaxy bispectrum, one needs to integrate our theoretically calculated bispectrum in \eqref{final} over the three window functions describing the redshift bins
\be 
\label{windowed}
b_{\ell_1,\ell_2,\ell_3}^W(z_1,z_2,z_3) = \int {\rm d} z_1' \ {\rm d}  z_2'  \ {\rm d} z_3' \  W(z_1,z_1')   W(z_2,z_2')  W(z_3,z_3') \ b_{\ell_1,\ell_2,\ell_3}(z'_1,z'_2,z'_3) \ .
\ee
Given our expression in \eqref{final}, it is straightforward to compute $b_{\ell_1,\ell_2,\ell_3}^W(z_1,z_2,z_3)$.


\subsection{The distant observer approximation}\label{sec:flat}
In order to build some intuition, let us rewrite our  expression \eqref{final} in the distant observer approximation where we choose to decompose in Fourier modes only the direction perpendicular to the line of sight,

\begin{multline}
\Big< \Delta_{\rm g}(z_1, \vec{\ell}_1) \Delta_{\rm g}(z_2, \vec{\ell}_2) \Delta_{ \rm g}(z_3,\vec{\ell}_3) \Big>_ {\ell_1 \ll \ell_2, \ell_3} = (2\pi)^2 \delta_{\rm D}(\vec{\ell}_1+\vec{\ell}_2+\vec{\ell}_3) \bigg[C^{\Delta_{\rm g} d}(z_1, z_2,\ell_1)
 \\
+ C^{\Delta_{\rm g} \Delta z}(z_1, z_2, \ell_1) \frac{\de}{\de z_2} + \vec{\ell}_1\cdot\vec{\ell}_3 C^{\Delta_{\rm g} I}(z_1, z_2, \ell_1) \bigg] C^{\Delta_{\rm g} \Delta_{\rm g}}(z_2, z_3, \ell_3) + (2 \leftrightarrow 3).
\end{multline}
Note that the third term in the square brackets,  when added to the corresponding one coming from $(2 \leftrightarrow 3)$,  is ${\cal O}(\ell_1^2)$ in the equal time limit as expected due to momentum conservation, while the other terms are different from zero in that limit. Note also that the Newtonian consistency relations take into account only the redshift space distortions induced by the long-wavelength velocity, which are included here in $\Delta z$. This Newtonian contribution is not zero in the equal-time limit. The same would hold if one also decomposes the direction parallel to the line of sight in Fourier modes. This is due to the fact that one performs measurements on the light-cone, and since there is no translation invariance in the radial direction in redshift space, the two momenta connected to this radial direction would not cancel. However, this non-cancellation will be governed by how much the power spectrum changes with redshift; when it is a good approximation to ignore the evolution of the power spectrum this Newtonian consistency relation is expected to be zero as was previously found. Indeed, in the above expression it is proportional to a derivative of the power spectrum with respect to redshift. Let us mention, however, that the contribution from this term is very small due to the fact that velocity and density are not correlated in harmonic space.

In the rest of this paper, we use the general expression given in Eq. \eqref{final} without resorting to the flat sky approximation. 

\subsection{Second-order limit}
\label{sec:second-order}

\begin{figure}[t]
\centering
\includegraphics[width=0.7\textwidth]{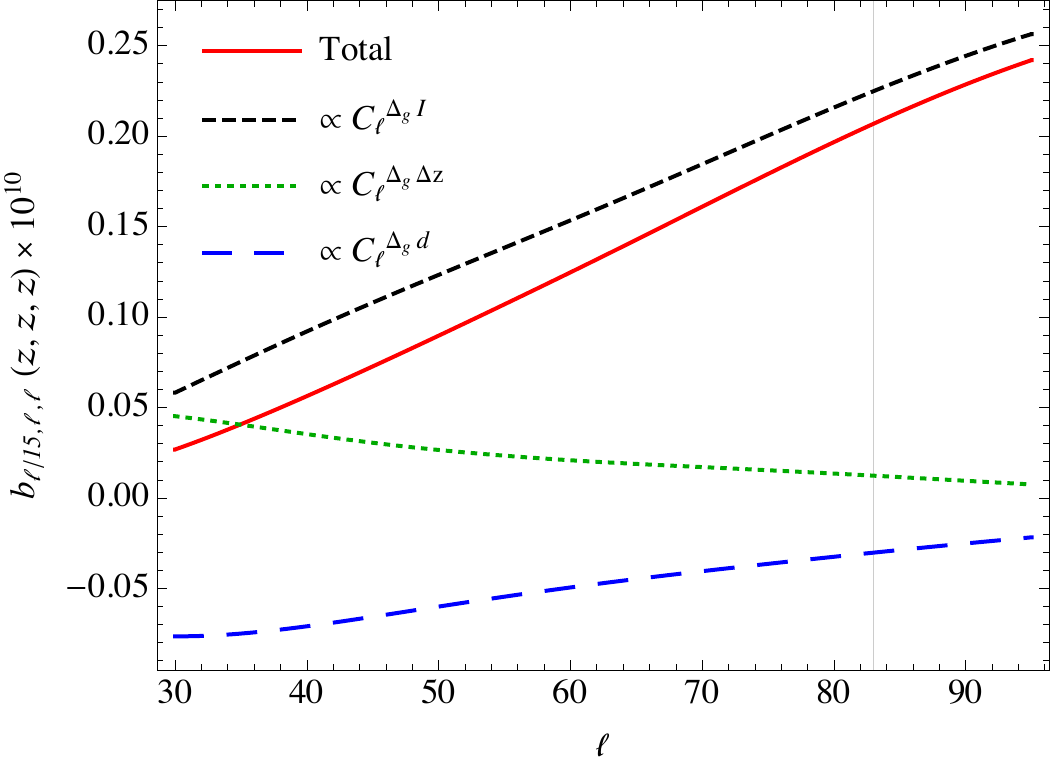}
\caption{\textit{Reduced observed bispectrum in the squeezed limit at equal-time (z=0.55) on linear scales. Note that the consistency relation is also valid when the short modes $\ell_2, \ell_3$ are in the non-linear regime. The vertical line indicates the linear scale up to which the numerical code is valid.}}
\label{f1}
\end{figure}

As an illustration, we plot in figure  \ref{f1} different terms contributing to the reduced bispectrum appearing in the relation \eqref{final} in the case where the short modes are linear. The second-order computation of the three-point correlator of the observed galaxies should reproduce this result in the squeezed limit. We do this using a modified version of the code CLASSgal \cite{DiDio:2013bqa}. The result is presented in the simple case of no galaxy bias $b=1$, no magnification bias $t=0$ and no galaxy evolution $f_{\rm evo}=0$ for equal redshifts of 0.55.
The vertical line denotes the linear scale at $z=0.55$ of $\ell\simeq 83$ up to which the numerical code is valid.

One could be surprised by the small size of the cross correlation between $\Delta_{\rm g}$ and $\Delta z$. In the Newtonian approximation and working at a fixed time slice, the terms proportional to $\Delta z$ are indeed believed to be the dominant effect. However, the term is suppressed in harmonic space due to the fact that the Bessel functions corresponding to the $\Delta_{\rm g}$ term and  the $\n \cdot \vec{v}_L$ term are $j_\ell$ and $j_\ell'$ which oscillate at a similar frequency but out of phase and  therefore cancel. In Fourier space, this can be intuitively understood from the fact that the regions where the velocities are at the maximum correspond to the minima (in absolute value) of the density and vice versa. Moreover the lensing contribution is smaller than one would expect parametrically. This is due to the fact that the integration along the line of sight tends to average out the perturbations.

\section{Effective local non-Gaussianity parameter from non-linear general relativity corrections}
\label{sec:fnl}

\begin{figure}[t]
\centering \hspace{0.1cm}
\includegraphics[width=0.68\textwidth]{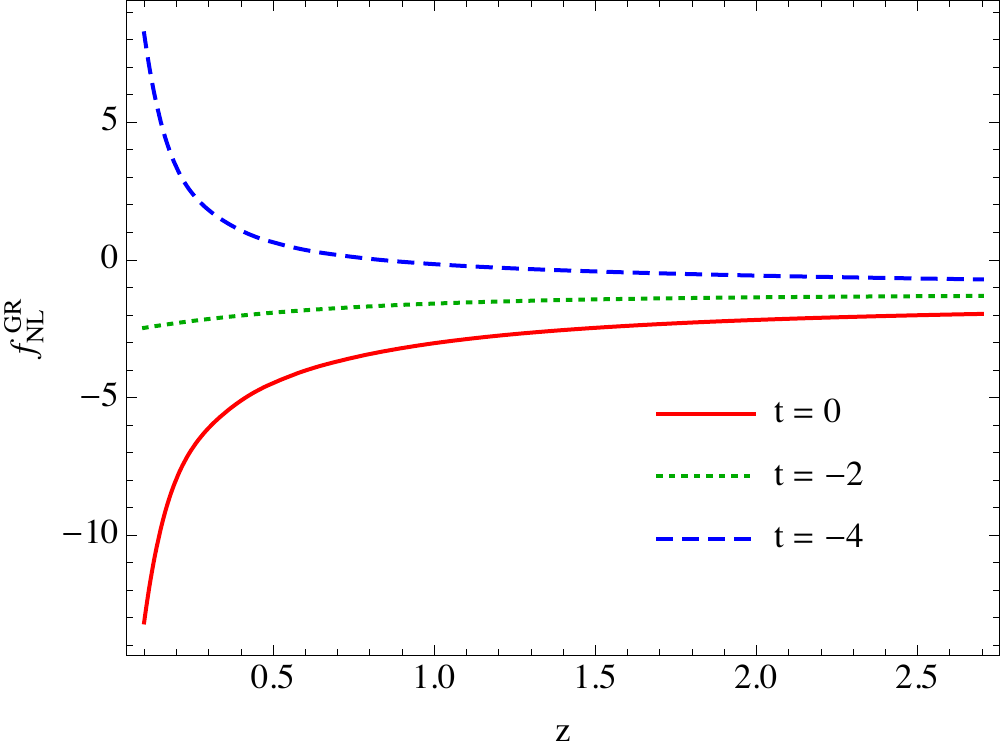}
\caption{\textit{Effective $f^{\rm GR}_{\rm NL}$ induced by the relativistic effects computed in this paper. The three different lines correspond to different magnification bias parameters. We have assumed $e = 3/a$ for all cases. This estimate is valid for $z\gtrsim 1$. At lower redshifts, integration over the redshift bins can change this result.}}
\label{fig:fnl}
\end{figure}

As mentioned in previous sections, the terms proportional to $\Phi \Delta_{\rm g}$ would induce a signal in the observations that might be misinterpreted as primordial  non-Gaussianity with a local shape. In order to see this, let us estimate the squeezed limit of the galaxy bispectrum induced by the primordial local non-Gaussianity. From a Newtonian calculation and assuming linear biasing we have\footnote{A similar expression is obtained in Ref. \cite{Peloso:2013zw} for the dark matter over-density bispectrum and it is expected to hold at the non-perturbative level. However, it is  not clear whether this is also true for galaxies.}
\be
\Big< \alm{1} \alm{2} \alm{3} \Big>_{\ell_1 \ll \ell_2, \ell_3} = 4 f_{\rm NL} \gaunt C^{\Delta_{\rm. g}\Phi}_{\ell_1}(z_1,z_2)C^{\Delta_{\rm g}\Delta_{\rm g}}_{\ell_3}(z_2,z_3)\,,
\label{eq:fnl}
\ee
where \hbox{$f_{\rm NL} \sim 1$} is the non-linear non-Gaussianity parameter parametrizing the strength of the primordial non-Gaussianity with a local shape. To see the impact of the non-linear relativistic contributions, we compute the ratio of Eq.~\eqref{final} to $C^{\Delta_{\rm g}\Phi}_{\ell_1}C^{\Delta_{\rm g}\Delta_{\rm g}}_{\ell_2}$ where the potentially highly non-linear short-mode power spectrum cancels such that our result is fully non-perturbative\footnote{This cancelation does not happen for the term proportional to $C^{\Delta_{\rm g}\Delta z}_{\ell_1} \partial C^{\Delta_{\rm g}\Delta_{\rm g}}_{\ell_2}/\partial z$. However, as one can easily check at second-order,  this term  has a different $\ell$-dependence than Eq.~\eqref{eq:fnl}. This is because the derivative with respect to redshift acting on the short mode power spectrum, acts on the Bessel function and will induce a different $\ell$-dependence. A similar $\ell$-dependence will  also arise at higher orders}. We then isolate the terms for which this ratio is independent of $\ell$, \emph{i.e.} those terms in Eq.~\eqref{final} with the same $\ell$-dependence as Eq.~\eqref{eq:fnl}. In order to compare with the observed bispectrum in a galaxy survey, one needs to integrate over the window functions defining the redshift bins before taking this ratio. Unless the window functions are delta functions, which for a realistic survey they are not, this integration will change both the amplitude and the $\ell$-dependance of the bispectrum. However, due to the nearly constant behaviour of the transfer functions at $z\gtrsim1$, the non-linear power spectrum of the short modes still cancels in the ratio. Therefore this integration does not modify our estimate of $f_{\rm NL}^{\rm GR}$ for those redshifts. For lower redshifts, one can explicitly carry out the integration, though the results would be only perturbative in the short modes. It should be noted that since the signal for the primordial non-Gaussianity is strongest at higher redshifts, removing the GR effects that mimic this signal is of crucial importance at those redshifts.

The exact value of $f_{\rm NL}^{\rm GR}$ will depend on redshift and the magnification bias.  We plot its value  as a function of redshift in figure \ref{fig:fnl} for three different values of the magnification bias. From these results we deduce that at redshift $z=1.5$ and for a vanishing magnification bias, the value of the  effective $f^{\rm GR}_{\rm NL}$ is about $- 3.0$.  A qualitatively similar result was obtained in \cite{Jeong:2011as} for the power spectrum. They found a very similar dependance on redshift and magnification bias as well as a value for $f^{\rm GR}_{\rm NL}$ of the same order of magnitude. Since they considered a different observable, the exact values are different though their behavior is qualitatively the same.

Also, these relativistic effects could also induce terms that would be mistaken for a violation of the Equivalence Principle in a naive analysis \cite{Creminelli:2013nua,Kehagias:2013rpa}. 

\section{Conclusions}\label{sec:conc}

In this paper we have exploited the fact that a long-wavelength mode of Bardeen's potential is an adiabatic mode to write a non-perturbative relativistic relation for the squeezed limit of the galaxy number overdensity three-point function. Adiabatic modes are physical large-scale solutions to the dynamics that can be cancelled by a residual gauge transformation. The latter is realized non-linearly by Bardeen's potential. Thus, computing their effect on short-scale physics reduces trivially to a change of frame. 

The effect of these transformations on the observed galaxy number density $n_{\rm g}^{\rm obs}$ corresponds to a change in the redshift and the direction of the photons at observation which in turn changes the relation between the observed volume and the physical volume, and a magnification bias. When computing the galaxy number overdensity $\Delta_{\rm g}$, one should also include the change of the average $\bar{n}_{\rm g}$ as a result of distortions to the redshift, the so-called evolution bias. 

Curvature and tidal contributions, proportional to second derivatives of the gravitational potential, are not captured by our argument and are expected to be larger inside the horizon than some of the relativistic corrections which we keep, even though they would naively be expected to be parametrically subdominant with respect to the redshift space distortions and lensing induced by the long mode. However, redshift space distortions, which were also obtained in a Newtonian computation, are small due to the requirement that the small-scale modes be on similar redshifts (though they will be non-zero even at equal redshift), and the lensing is  suppressed since it averages in the trajectory of the photon from emission to observation.  We argue that the relativistic corrections we compute can however be distinguished by their dependence on the large scale as one takes the squeezed limit, and they serve as a consistency check for our cosmological model. Indeed, the violation of the Equivalence Principle at large scales or the presence of primordial local non-Gaussianity would induce a squeezed limit with the shape of these contributions. They need to be kept in order not to confuse them with these deviations from the vanilla cosmological scenario. The exact value of the effective local non-Gaussianity parameter from these
non-linear GR corrections depends on the redshift and the magnification bias.  At redshift of $z=1.5$ and in the absence of magnification bias we estimate \hbox {$f^{\rm GR}_{\rm NL} = - 3.0 $}. 

Finally, since the universe evolves and there is no translation invariance in the radial direction of redshift space, Fourier-space triangles do not close and we thus present our expressions after decomposing them in spherical harmonics. Notably, we find that the Newtonian redshift space distortions induce a term that does not go to zero at equal redshifts, but is however very small.

There are many directions in which this work can be extended. It would be interesting to see to which level of accuracy planned large scale structure surveys could put constraints on these consistency relations. The prospect is promising since one can use arbitrarily small scales for the short modes given that the consistency relations are non-perturbative. However, the effect of systematic observational effects must be taken into account. Another interesting direction would be to perform a similar calculation for the galaxy two-point function including contributions from the dark matter bispectrum, thus affecting the measured bias and its scale dependence which is sensitive to the squeezed limit; such computation would include all the relativistic and non-linear effects.

\section*{Acknowledgements}
We would like to thank R. Durrer, H. Gil-Marin an M. Manera for interesting discussions and E. Di Dio and F. Montanari for the help in using the CLASSgal code. The research of A.K. is implemented under the Aristeia II Action of the Operational Program Education and
Lifelong Learning and is co-funded by the European Social Fund (ESF) and National
Resources. A.K. is also partially supported by European Unions Seventh Framework
Program (FP7/2007-2013) under REA grant agreement n. 329083. H.P. , J.N. and A.R. are supported
by the Swiss National Science Foundation (SNSF), project ``The non-Gaussian
Universe" (project number: 200021140236). A. M. is supported by the Tomalla foundation for Gravity Research.

\bibliography{SCINC}

\end{document}